\documentclass[epjc3]{svjour3}
\journalname{Eur. Phys. J. C}
\RequirePackage{graphicx}
\usepackage[utf8]{inputenc}
\usepackage{hyperref}
\usepackage{subcaption}
\usepackage{upgreek}
\usepackage{siunitx}
\usepackage{booktabs}
\usepackage[normalem]{ulem}
\usepackage{mhchem}
\usepackage[acronym]{glossaries}
\usepackage[dvipsnames]{xcolor}
\newacronym{LDLS}{LDLS}{Laser-Driven Light Source}
\usepackage[final]{changes}
\usepackage{comment}
\usepackage{lineno}
\graphicspath{{figures/}}

\newcommand{\CW}[1]{\textcolor{black}{#1}} 
\newcommand{\CWCW}[1]{\textcolor{black}{#1}} 

\newcommand{\Sch}[1]{\textcolor{black}{#1}} 

\begin{document}
\title{A precision 32$\,$keV angular-selective photoelectron source for calibration measurements at the KATRIN experiment}

\titlerunning{\parbox{9cm}{A precision high-energy photoelectron source}} 
\authorrunning{S.~Schneidewind, \dots, C.~Weinheimer}
\institute{%
Institute for Nuclear Physics, University of Münster, Wilhelm-Klemm-Str. 9, 48149 M\"{u}nster, Germany\label{a}
\and Istituto Nazionale di Fisica Nucleare (INFN)-Sezione di Milano-Bicocca, Piazza della Scienza 3, 20126 Milan, Italy\label{b}
\and Institute for Astroparticle Physics (IAP), Karlsruhe Institute of Technology (KIT), Hermann-von-Helmholtz-Platz 1, 76344 Eggenstein-Leopoldshafen, Germany\label{c}
\and Center for Experimental Nuclear Physics and Astrophysics, and Dept. of Physics, University of Washington, Seattle, WA 98195, USA \label{d}
\and Max-Planck-Institut für Kernphysik, Saupfercheckweg 1, 69117 Heidelberg, Germany\label{e}
\and Institute of Experimental Particle Physics (ETP), Karlsruhe Institute of Technology (KIT), Wolfgang-Gaede-Str. 1, 76131 Karlsruhe, Germany\label{f}
\and Institute for Data Processing and Electronics, Karlsruhe Institute of Technology, Hermann-von-Helmholtz-Platz 1
D-76344 Eggenstein-Leopoldshafen, Germany\label{g}
}
\thankstext{email1}{e-mail: sonja.schneidewind@mib.infn.it}
\thankstext{email2}{e-mail: hannen@uni-muenster.de}
\thankstext{email3}{e-mail: weinheimer@uni-muenster.de}
\author{%
S.~Schneidewind\thanksref{a,b,email1}
\and R.~Sack\thanksref{c}
\and F.~Block\thanksref{c}
\and S.~Enomoto\thanksref{d}
\and V.~Hannen\thanksref{a,email2}
\and C.~Köhler\thanksref{e}
\and A.~Lokhov\thanksref{f}
\and A.~Marsteller\thanksref{c}
\and H.-W.~Ortjohann\thanksref{a}
\and R.W.J.~Salomon\thanksref{a}
\and L.~Schimpf\thanksref{c}
\and K.~Schlösser\thanksref{c}
\and S.~Wüstling\thanksref{g}
\and C.~Weinheimer\thanksref{a,email3}
}

\date{\today}
\maketitle

\begin{abstract}
The Karlsruhe Tritium Neutrino (KATRIN) experiment measures the neutrino mass from a precise measurement of the endpoint region of the kinematic tritium $\upbeta$-decay spectrum by using a spectrometer combining magnetic adiabatic collimation and electrostatic filtering (MAC-E filter). For calibration purposes, KATRIN uses a monoenergetic angular-selective photoelectron source. We present an upgrade of this source, which was installed in the KATRIN beamline in February 2022. The source allows for a wide range of accessible electron energies up to \SI{32}{\kilo\electronvolt} and a variation of the angle with regard to the magnetic field. These features are used for precise measurements of electron scattering effects off tritium molecules in \mbox{KATRIN}'s gaseous tritium source, for investigations of angular-dependent backscattering for example at \mbox{KATRIN}'s focal-plane detector, and for studies on adiabatic transport in the main spectrometer.
\end{abstract}

\section{Introduction}
\label{sec:intro}
At the Karlsruhe Tritium Neutrino (KATRIN) experiment\,\cite{KATRIN_2005}, the neutrino mass is measured directly from the kinematics of tritium $\upbeta$ decay, currently setting the world-best upper limit of \SI{0.45}{\electronvolt\per c^2} (\SI{90}{\percent} C.L.) on the effective electron antineutrino mass\,\cite{KATRIN:2024cdt}. \mbox{KATRIN} combines a windowless gaseous molecular tritium source with the technique of magnetic collimation and electrostatic filtering in a MAC-E filter type spectrometer with negative retarding voltage. The kinematic integral $\upbeta$-decay spectrum is measured close to its endpoint at \SI{18.6}{\kilo\electronvolt} by counting
the number of electrons transmitted at different retarding voltage set points. The measured spectrum is described by a convolution of the differential energy spectrum with the experimental response of \mbox{KATRIN} plus a constant background term. 
First, the response includes the transmission properties of the spectrometer, which rely on the knowledge of the electric and magnetic fields. Second, it contains information on energy losses due to scattering of electrons with tritium molecules in the source. The experimental response is determined by means of the calibration sources, namely a gaseous $^{\text{83m}}$Kr source and a pulsable monoenergetic, angular-selective photoelectron source\,\cite{Aker_2021_hardware,KATRIN:2025tbl}. 

In this work, an upgrade of the photoelectron source, installed in the \mbox{KATRIN} beamline in February 2022, is presented. Furthermore, improved methods for the characterization of the properties of the electron source, especially regarding the angle of the emitted photoelectrons relative to the magnetic-field lines, are introduced. The paper is structured as follows: Sec. \ref{sec:egunprinciple} introduces the general working principle of the photoelectron source. The environment in which the electron source is installed at \mbox{KATRIN}, the so-called Rear Section at the beginning of the \mbox{KATRIN} beamline, is described in sec. \ref{sec:rearsection}, followed by a description of the upgraded electron-source design in sec. \ref{sec:newdesign}. Details on the experimental response of the photoelectron source are given in sec. \ref{sec:expresponse}. A new method to estimate the average electron angle from measurements of the transmission function at different magnetic-field settings
is introduced in sec. \ref{sec:pitchangle}. Finally, a novel hardware-based method for a reduction of the background level of the photoelectron source is presented in sec. \ref{sec:background}.

\section{Principle of the photoelectron source\label{sec:egunprinciple}}

\begin{figure}
    \centering
    \includegraphics[width=0.6\linewidth]{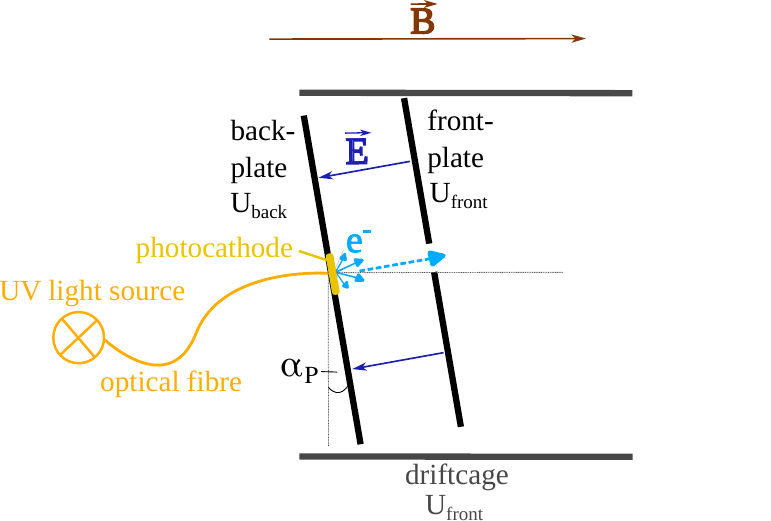}
    \caption{Schematic working principle of the photoelectron source. UV light is guided via an optical fibre towards the photocathode, where low-energy electrons are emitted. The cathode is placed on the backplate and put at a negative high voltage, so the electrons are accelerated by the electric field $\vec{E}$ between back- and frontplate. 
    The electron angle can be altered by adjusting the plate angle $\alpha_\text{P}$ with regard to the magnetic field $\vec{B}$.}
    \label{fig:egunschema}

\end{figure}

The basic principle of the photoelectron source was introduced in \cite{Valerius_2011} and worked out in more detail in \cite{Beck_2014} and \cite{Behrens2017}. It is based on the creation of photoelectrons via ultraviolet (UV) back-illumination of a gold or silver photocathode and the acceleration of emitted electrons in an electric field $\vec{E}$ created by two parallel electrodes at different potential. The electron pitch angle $\theta$ relative to the local magnetic field $\vec{B}$ can be adjusted via a tilt of the two electrodes by an angle $\alpha_\text{P}$ relative to the magnetic-field lines, as illustrated in fig. \ref{fig:egunschema}. 

In a simplified picture, photoelectrons can be emitted if the photon energy $E_\uplambda$ for light with wavelength $\lambda$ is higher than the work function $\Phi$ of the cathode, resulting in a positive electron energy $E_\text{e}$:
\begin{equation}
E_\text{e} = E_\uplambda - \Phi = \frac{h\cdot c}{\lambda} - \Phi >0\,. \label{equ:Ee}
\end{equation}
The electron source presented in this work has a gold layer of $\mathcal{O}(\SI{20}{\nano\meter})$ thickness, evaporated onto the front face of a UV transparent optical fiber. The layer acts as a photocathode, which can be illuminated by two different UV light sources.
In the current \mbox{KATRIN} setup, electron emission is possible using light wavelengths below $\approx\SI{285}{\nano\meter}$. A typical value is $\lambda = \SI{266}{\nano\meter}$, corresponding to a photon energy of $E_\lambda = \SI{4.7}{\electronvolt}$. The work function of gold mentioned in literature is $\qtyrange[range-units=bracket]{4.7}{5.4}{\electronvolt}$, but it is strongly dependent on the crystal orientation as well as on parameters like purity, surface roughness, grain size and vacuum conditions \cite{Anderson_1959,Sachtler_1966,Riviere_1966,Aghili_2024}. 

The width of the energy distribution of photoelectrons is a key parameter for the photoelectron source. The energy distribution of emitted electrons becomes broader for light with shorter wavelengths, corresponding to larger $E_\lambda$, and can be minimized by choosing $E_\lambda$ just slightly above the photocathode's work function $\Phi$. For a metal, the work function can be identified with the Fermi energy, $\Phi=E_\text{F}$, and it describes the electron state occupation probability in the \replaced{conduction}{conductance} band by the Fermi 
distribution:
\begin{equation}
f(E, T) = \frac{1}{\exp{(\frac{E-E_\text{F}}{k_\text{B}T})}+1}\,.
\end{equation} 
Therefore, the difference between photon energy and work function $E_\lambda - E_\text{F}$ as well as a broadening related to the (effective) temperature $T$ determine the energy distribution of emitted electrons. More precisely, the fraction of electrons in the \replaced{conduction}{conductance} band with a binding energy smaller than the photon thres\-hold broadens the photoelectron energy distribution at temperatures $T>\SI{0}{\kelvin}$. The broadening can amount up to about \SI{100}{\milli\electronvolt} at $T = \SI{300}{\kelvin}$. In addition, a spread of the work function (Fermi energy) due to inhomogeneities or impurities over the photocathode, as well as the spectral width of the light source will further broaden the energy distribution of the emitted photoelectrons. In general, a smaller photon energy $E_\lambda$, even below the mean work function $\langle \Phi \rangle = \langle E_\text{F} \rangle$, will -- at the cost of a lower count rate -- reduce the width of the photoelectron energy spectrum.\\

After being emitted from the photocathode at potential $U_\text{back}$, the electrons are \Sch{accelerated} strongly non-adiabatically by the Lorentz force $\vec{F}_\text{L}$ 
via the electric field $\vec{E}$ on the order of $\SI{3e5}{\volt\per\meter}$ created by the voltage difference $q(U_\text{front} - U_\text{back})$ between front- and backplate, and $\vec{B}$ with $\vert \vec{B} \vert\approx \SI{25}{\milli\tesla}$ for the present source:

\begin{equation}
\vec F_\text{L} = q\cdot (\vec E + \vec v \times \vec B)\,. \label{equ:lorentz}
\end{equation}
Here, $q=-e$ is the electron charge and $\vec{v}$ the electron velocity. After emission, the electrons are firstly accelerated parallel to $\CW{q}\vec{E}$. With increasing transversal velocity $v_\perp$ (perpendicular to $\vec{B}$), the magnetic term in eq. \ref{equ:lorentz} becomes more important. 
When the electric field becomes zero, the electron movement is dominated by adiabatic transport in which the magnetic orbital momentum

\begin{equation} 
\mu_\text{\added{l}} = \frac{1}{2m_\text{e}} \frac{p_\perp^2 }{\vert \vec{B} \vert} = \frac{1}{2m_\text{e}} \frac{p_\perp^2}{B} \label{equ:adiabatic_invariance}
\end{equation}
is conserved, with the electron momentum perpendicular to the magnetic field direction
\begin{equation}
p_\perp = \gamma \cdot m_\text{e} \cdot v_\perp = \gamma \cdot m_\text{e} \cdot v\cdot \sin{\theta}\,,
\end{equation}
with electron mass $m_\text{e}$ and Lorentz factor $\gamma$. For the case $v=const. \Leftrightarrow \gamma = const.$ it follows from eq. \ref{equ:adiabatic_invariance} that $\sin^2{(\theta)}/B=const.$ and, thus, for the evolution of the pitch angle $\theta$ caused by the change of the magnetic-field strength $B=\vert \vec{B} \vert$

\begin{equation} \label{eq:adiabatic_transport}
    \frac{B_1}{B_2} = \frac{\sin^2{(\theta_1})}{\sin^2{(\theta_2)}}\, \rightarrow \theta_\text{2} = \arcsin{\left( \sqrt{\frac{B_\text{2}}{B_\text{1}}}\cdot \sin{(\theta_\text{1}) }\right)}.
\end{equation}
The adiabatic invariant from eq. \ref{equ:adiabatic_invariance} can also be expressed in terms of the \replaced{transverse}{transversal} kinetic energy $E_\perp = E_\text{kin} \cdot \sin^2{(\theta})$ by using 
$\gamma^2 \cdot \beta^2 = \gamma^2 - 1 = (\gamma-1) \cdot (\gamma+1)$ and $E_\text{kin} = (\gamma -1) \cdot m_\text{e}c^2$\,\cite{Kleesiek:2018mel}:
\begin{equation}
    \mu_\text{\added{l}} = \frac{1}{2m_\text{e}} \frac{p_\perp^2 }{B} 
     = \frac{\gamma^2 \cdot \beta^2 \cdot m_\text{e} c^2 
     \cdot \sin^2{(\theta)}}{2 \cdot B}
     = \frac{(\gamma+1) \cdot E_\text{kin} \cdot \sin^2{(\theta)}}{2 \cdot B}
     = \frac{(\gamma+1) \cdot E_\perp}{2 \cdot B}\,,
     \label{equ:relativistic_adiabatic_invariant}
\end{equation}
which transforms for $\gamma \approx 1$ into the well-known non-relativistic invariant 
\begin{equation}
\mu_\text{\added{l}} = \frac{E_\perp}{B} = const.
\end{equation}

\section{\Sch{Setup at the KATRIN Rear Section} \label{sec:rearsection}}
\begin{figure}
    \centering
    \includegraphics[width=0.8\linewidth]{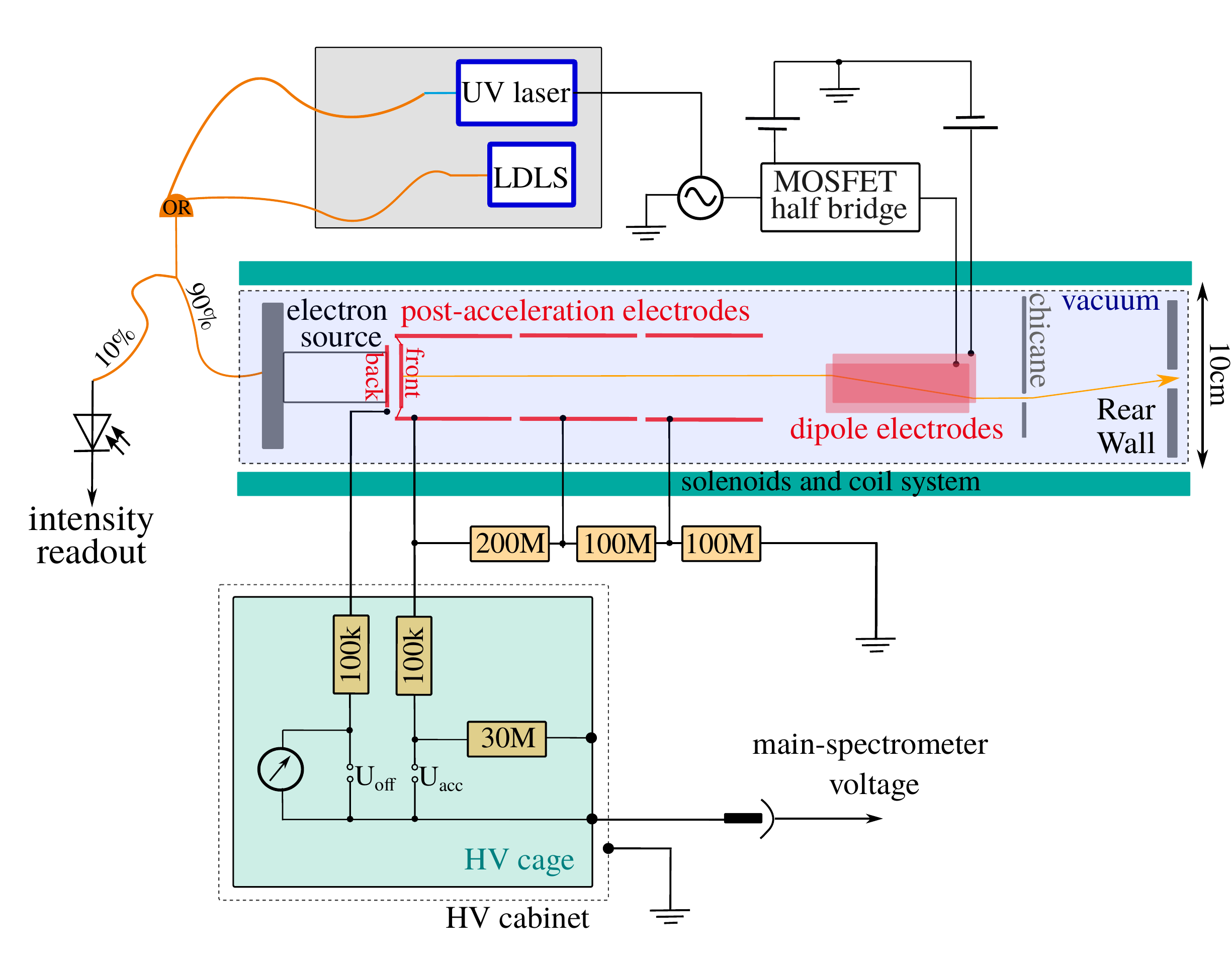}
    \caption{Illustration of the electron-source setup at the \mbox{KATRIN} Rear Section. The backplate voltage is defined by the voltage of the HV cage and an offset voltage $U_\text{off}$ within     $\pm \SI{500}{\volt}$. The frontplate voltage is provided by an additional high voltage $U_\text{acc}$. 
    The electric potential is stepwise reduced to ground via three post-acceleration electrodes connected to a HV divider. The acceleration takes place in the first \SI{0.5}{\meter} of the Rear Section with a total length of around \SI{5}{\meter}. The electron beam is steered by several magnetic dipole coils (not shown) and a set of dipole electrodes to pass the Rear-Wall hole and the chicane towards the beamline. The chicane prevents damage from the electron source by \replaced{ensuring}{avoiding} that neutral molecules from the tritium source \replaced{cannot}{can} directly travel towards the sensitive photocathode of the electron source. One of the dipole electrodes can be pulsed via a MOSFET half bridge for background reduction, see sec. \ref{sec:background}. The guiding magnetic field in the Rear Section is provided by several solenoids. }
    \label{fig:rearsection}
\end{figure}

At \mbox{KATRIN}, the photoelectron source is located in the so-called Rear Section\,\cite{Aker_2021_hardware,Babutzka_2014}, which is the rear end of the beamline, the furthest away from the spectrometer. From there, emitted electrons pass the complete \SI{70}{\meter} long beamline on their way to the spectrometer and the detector, guided by adiabatic transport. The environment in which the electron source is installed is schematically shown in fig. \ref{fig:rearsection}.
 A driftcage around front- and backplate at potential $U_\text{front}$ improves field homogeneity. To avoid abrupt potential changes between driftcage and ground potential, two additional cylindrical post-acceleration electrodes at certain fractions of $U_\text{front}$ are installed to smoothly accelerate the electrons to their final energy $E_\text{kin} = q U_\text{back}$. The fractional voltages are realized using a voltage divider with resistance ratio 2:1:1. 
 
 The pitch angle $\theta$ \replaced{with respect to the}{w.r.t.} magnetic field at the photoelectron source depends on the parameters $\vert\vec{E}\vert$, $\vert\vec{B}\vert$ and $\alpha_\text{P}$.
After passing the post-acceleration electrodes, the electrons perform an adiabatic cyclotron motion along the magnetic-field lines. Two sets of magnetic dipole coils, internally called steering and compensation coils, are used to guide the electrons through a $\SI{5}{\milli\meter}$ diameter hole in the Rear Wall towards the gaseous tritium source. The Rear Wall is a gold-plated stainless steel disk which terminates the $\beta$-electron flux in backwards direction, and defines the reference potential of the tritium source. 

\added{If tritium molecules decay between front- and backplate of the electron source, the remaining ions are accelerated towards the sensitive photocathode and produce an undesired background rate (see sec. \ref{sec:background}), or even destroy the cathode, if there are too many decays.} To \replaced{prevent}{avoid that} neutral molecules from the tritium source \replaced{from reaching the electron source}{reach the sensitive photocathode surface}, the direct line of sight between photocathode and tritium source is \replaced{blocked}{prevented} by a chicane \replaced{with an aperture}{of which the aperture has a design offset} of \SI{10}{\milli\meter} \replaced{relative}{with regard} to the Rear-Wall hole. The electron beam is steered with the dipole coils around the chicane. 
To remove stored charged particles\,\footnote{\added{tritium ions, $\upbeta$-decay electrons or electrons created for example by field emission at the photoelectron source or the post-acceleration electrodes}} from the Rear Section, the two dipole electrodes make use of the $E\times B$ drift inducing a drift velocity\,\cite{Jackson_2014}
\begin{equation}
    \vec{v}_\text{D} = \frac{\vec E \times \vec{B}}{\vert\vec{B}\vert^2}\,.
\end{equation}

The power supplies for the different voltages are located in a high-voltage (HV) cage. 
The cage is typically coupled to the main-spectrometer vessel voltage $U_\text{vessel}$, in order to profit from the excellent HV stabilization system of the main spectrometer\,\cite{Rodenbeck:2022iys}. The potential
$U_\text{back}$ is defined by the HV cage voltage plus a voltage offset $U_\text{off}$ within $\pm \SI{500}{\volt}$ provided by an 
ISEG\footnote{iseg Spezialelektronik GmbH, Bautzner Landstr. 23, 01454 Radeberg, Germany} MMC-500V power supply to enable \Sch{a} variation of the surplus energy of electrons relative to the retarding potential at the spectrometer. It is measured precisely via a 6.5-digit Fluke\footnote{Fluke Deutschland GmbH, Hardstraße 20, 8303 Bassersdorf, Switzerland} Digital voltmeter. The voltage between front and back plate $U_\text{acc}$ is the sum of $U_\text{off}$ and an additional voltage difference  of several kV which is realized via an ISEG \SI{15}{\kilo\volt} power supply.

The electron source features two basic operation modes: The first mode is a continuous mode which uses a Laser-Driven Light Source (LDLS) from Energetiq\,\footnote{Energetiq Technology, Inc., 205 Lowell Street, Wilmington, MA 01887, USA} \Sch{type} EQ-99X-FC for UV-light creation. The second mode is a pulsed mode using a pulsed UV-laser from InnoLas\,\footnote{InnoLas Laser GmbH, Justus-von-Liebig-Ring 8, 82152 Krailling} type Mosquitoo 266, which is a frequency quadrupled Nd:YVO$_\text{4}$ laser emitting a wavelength of \SI{266}{\nano\meter}.

The LDLS emits light with a wavelength spectrum between \SI{190}{\nano\meter} and \SI{2500}{\nano\meter}, and a monochromator of type SpectralProducts\,\footnote{Spectral Products, 111 Highland Drive, Putnam, CT 06260} DK 240 allows for wavelength selection with around \SI{6}{\nano\meter} resolution. The possibility to select a certain wavelength is useful for minimizing the energy spread of the electrons by choosing an optimal difference $E_\lambda - \Phi$, or for measurements to determine the work function of the photocathode. Typically, wavelengths between \SI{260}{\nano\meter} and \SI{285}{\nano\meter} are used. \\
The operation with the pulsed UV-laser enables measurements of the flight time of electrons, allowing to extract a differential energy spectrum from the otherwise integral energy measurement of the \mbox{KATRIN} main spectrometer\,\cite{KATRIN:2021rqj}. The laser can be operated with an external trigger signal, provided by a frequency generator of type RIGOL\,\footnote{Rigol Technologies EU GmbH, Carl-Benz-Strasse 11, 82205 Gilching, Germany} DG832. The typical operation frequency is \SI{20}{\kilo\hertz}, which is the minimal operation frequency of the laser. The laser intensity and therefore the electron rate can be adjusted via \CW{rotating} a $\lambda/4$ plate between the two frequency doubling stations \added{which quarter the laser wavelength from the initially emitted \SI{1024}{\nano\meter} to \SI{266}{\nano\meter}}. 

The light of both sources is coupled into optical fibres, of which one or the other can be connected to the electron source. A beamsplitter between light source and electron source \replaced{enables the monitoring of}{allows to monitor} the light intensity $I_\lambda$ in-situ via a Si-PIN photodiode of type 
 Thorlabs\,\footnote{Thorlabs GmbH, Münchner Weg 1, 85232 Bergkirchen,  Germany} SM1PD2A, whose photocurrent is read out via a 
 Femto\,\footnote{FEMTO Messtechnik GmbH, Klosterstraße 64, 10179 Berlin} DLPCA 200 transimpedance amplifier. Temperature stabilization is provided by an Arroyo\,\footnote{Arroyo Instruments, LLC, 1201 Prospect Street, San Luis Obispo, CA 93401, USA} 5300 TECSource for the LDLS and via a water chiller and an Arroyo 5400 TECSource for the laser.

\section{Design of the upgraded photoelectron source \label{sec:newdesign}}
Before the start of \mbox{KATRIN}'s operation in 2019, an angular-selective photoelectron source of a similar kind was temporarily installed close to the main spectrometer for spectrometer calibration purposes. This source, described in detail in \cite{Behrens2017}, allowed a tilt of the plates along two axes, realized by a combination of a tilt and a rotation mechanism. A second, angular-selective permanent photoelectron source 
was installed in the \mbox{KATRIN} Rear Section, allowing for the regular measurement of the gas density in the tritium source and the investigation of energy losses due to scattering of electrons off the molecules in the source \cite{Aker_2021_hardware,KATRIN:2021rqj}. This Rear-Section source allows a tilt in one direction only and thereby has a simpler mechanical design. In February 2022, an upgraded version of this source -- which is the topic of this paper -- was installed in the Rear Section, it has the following advantages compared to its predecessor:
\begin{itemize}
    \item an extended energy range, up to $\SI{32}{keV}$ instead of  $\SI{20}{keV}$, with an energy spread of $\mathcal{O}(\SI{100}{\milli\electronvolt})$ for any energy. This enables calibration measurements at the energies of the $N_\text{2,3}-32$ conversion lines of $^{\text{83m}}$Kr, which are used for investigation of the plasma properties of the tritium source\,\cite{KATRIN:2025tbl}, and for calibration of the electromagnetic fields in the spectrometer\,\cite{KATRIN:2024qct}\,;
    \item a precise and reproducible adjustment of the plate angle $\alpha_\text{P}$ with \SI{0.1}{\degree} precision, using a stepper motor (VAb\footnote{VAb Vakuum-Anlagenbau GmbH, Marie-Curie-Straße 11, 25337 Elmshorn, Germany} 16-25) \Sch{with a linear potentiometer for absolute position readout}\,;
    \item the pivot point of the tilting is changed from the frontplate to the backplate, such that the electron-emission point stays constant for any tilt angle. To facilitate this, the aperture of the frontplate was changed from a circular hole into a slit-shaped opening\added{, such that the electrons can pass the frontplate at all tilt angles}\,;
    \item an increased electron yield of $\mathcal{O}(\SI{20}{kcps})$ instead of $\mathcal{O}(\SI{1}{kcps})$ (using the same light source), achieved by a direct coupling of a single optical fiber to the photocathode instead of an indirect coupling of multiple fibers.
\end{itemize}
The crucial aspect of the upgraded design was to achieve the necessary HV stability of the source, as an electron energy of \SI{32}{keV} requires a voltage of \SI{-32}{kV} on the backplate. However, due to constraints from the Rear-Section environment\footnote{more precisely, voltage constraints for the high-voltage feedthroughs providing the voltages at the post-acceleration electrodes}, a maximum absolute voltage of \SI{20}{kV} is allowed on the frontplate. Consequently, voltage differences of at least $\SI{12}{kV}$ between front- and backplate are necessary. These requirements need to be fulfilled in the typical \mbox{KATRIN} operation conditions at a magnetic-field strength of $\mathcal{O}(\SI{25}{\milli\tesla})$, in vacuum conditions with a residual-gas pressure up to around \SI{1e-8}{\milli\bar}, and while photoelectrons are emitted from the photocathode. An
additional boundary condition is that the electron source needs to fit into the Rear Section
without additional modifications. Consequently, the frontplate needs to be constructed such that it matches the recess of the first post-acceleration electrode, and the whole source needs to fit into a DN160 tube. Furthermore, all parts have to be tritium-compatible, which mainly requires that they are free of halogen and hydrocarbons, as both substances can form corrosive compounds with tritium.

The breakdown voltage of electrodes strongly depends on electrode geometry, field homogeneity and residual-gas composition \cite{Lech2023}. An approximate formula describing the empirical relation between breakdown
 voltage $U_\text{break}$, gas pressure $p$ and the distance $d$ between two electrodes is Paschen's law\,\cite{Paschen1889,Townsend1915,Lieberman2005}:
 \begin{equation}
    U_\text{break} = \frac{B\cdot p \cdot d}{\ln{(A\cdot p \cdot d)} - \ln{\left[\ln{(1+\frac{1}{\gamma_\text{se}})}\right]}}.
 \end{equation}
 $A$ and $B$ are experimentally determined constants, and $\gamma_\text{se}$ is a coefficient describing the secondary-electron
 emission. Typically, around \SI{1}{\kilo\volt\per\milli\meter} can be applied between two electrodes. However, the magnetic field strength and direction, as well as a rough surface or sharp edges, harm the actual HV stability of a system. \CW{For this reason, all edges were rounded off and all stainless-steel parts were electropolished\,\cite{BastaniNejad_2015} to reduce surface roughness and thereby minimize field emission.}
 In addition, HV conditioning \cite{Guy_1973} can strongly enhance the HV capability of a system. Therefore, intensive HV tests and conditioning needed to be performed before the source could be installed in the \mbox{KATRIN} beamline.\\
 
\begin{figure}
    \centering
    \includegraphics[width=1\linewidth]{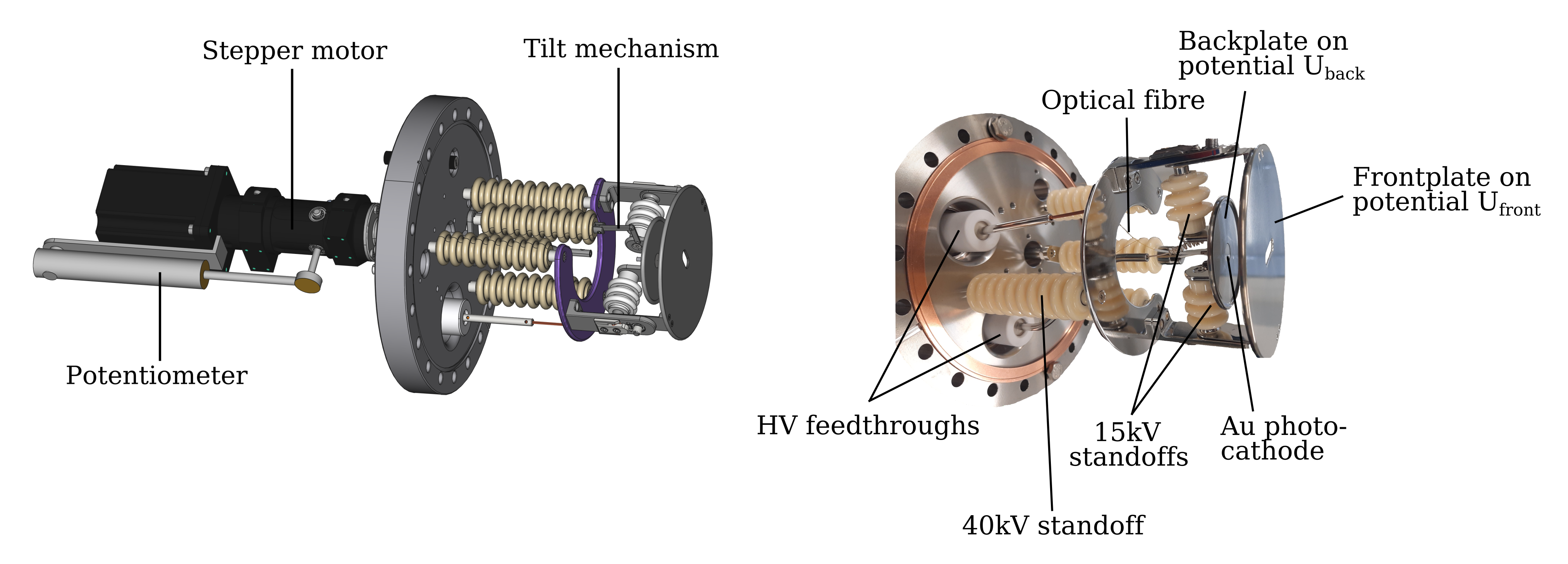}
    \caption{CAD drawing (left) and picture (right) of the upgraded photoelectron source. The source has a total length of \SI{33}{cm} (including the motor), and is hosted on a CF160 flange. \added{Motor and potentiometer are not visible in the picture on the right.}}
    \label{fig:cad}
\end{figure}

 The design of the upgraded source is shown in fig. \ref{fig:cad}. The two plates have a distance of \SI{2}{\centi\meter}. The frontplate has a diameter of \SI{10}{\centi\meter} and contains a slotted hole through which electrons can leave the electron source at all tilt angles up to $\alpha_\text{P}\approx \SI{14}{\degree}$. The backplate has a reduced diameter of \SI{6}{\centi\meter} to increase the distance to the side mounts on frontplate potential. To retain a homogeneous electric field despite the smaller diameter, the backplate rim is thicker than the center. The HV stability is supported by two \SI{15}{kV} type A standoffs from Kyocera\footnote{Kyocera Fineceramics Solutions GmbH, Steinzeugstraße 92, 68229 Mannheim, Germany} made of $\upalpha$-\ce{Al_2O_3} mounted below the backplate. \SI{50}{kV} HV feedthroughs and four Kyocera type B standoffs, designed for voltages up to \SI{40}{kV}, are used.
 
\begin{figure}
    \centering
    \includegraphics[width=0.6\linewidth]{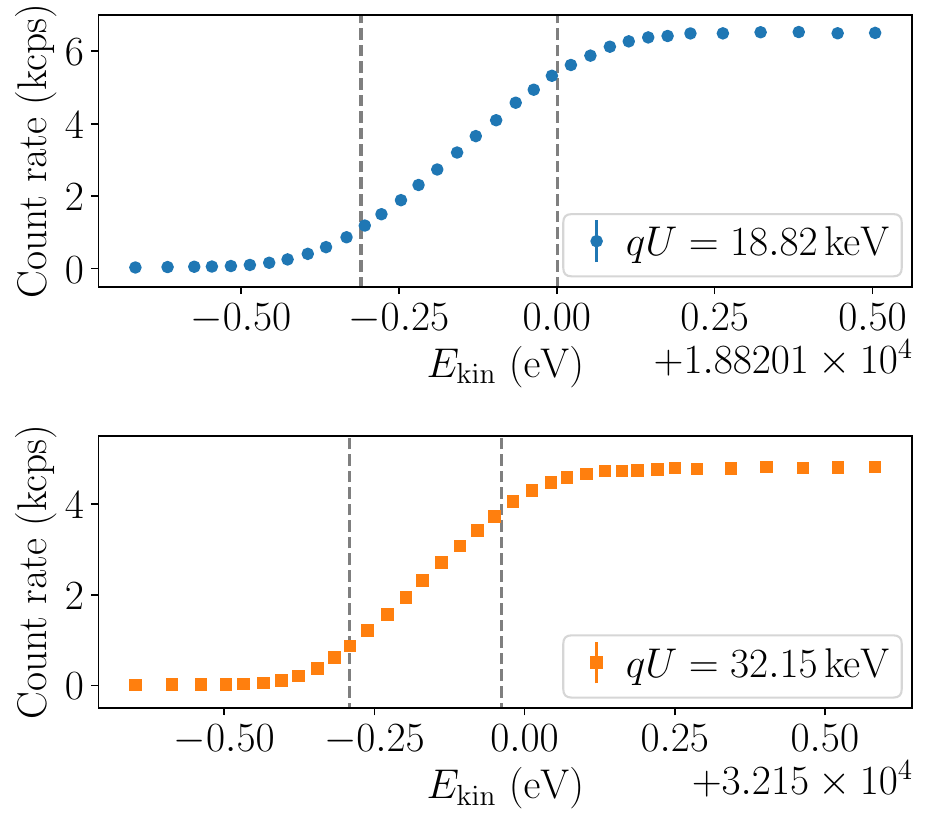}
    \caption{Transmission function of the photoelectron source measured at a standard energy of \SI{18.8}{\kilo\electronvolt} (top) and an enhanced energy of \SI{32.2}{\kilo\electronvolt} (bottom). Both graphs show the electron rate measured at the detector as function of electron energy $E_\text{kin} = qU_\text{back}$, with the spectrometer set to the retarding voltage $qU$ indicated by the legend. The dashed lines show the $1\sigma$ width of the functions. The maximum visible rate is around \SI{25}{\percent} of the total electron rate, because the measurements were taken while the beamtube was filled with tritium gas, such that $\approx \SI{75}{\percent}$ of the electrons lost energy due to inelastic scattering off tritium molecules and are not observed in the shown 
    energy range.}
    \label{fig:fpd-spectra}
\end{figure}

With this new design, a stable operation of the electron source at energies up to \SI{32}{\kilo\electronvolt} was achieved. This can be seen in fig. \ref{fig:fpd-spectra}, which shows two exemplary transmission functions measured at moderate energy and at \SI{32.2}{\kilo\electronvolt}. A transmission function is the transmitted electron rate as function of the difference between the longitudinal kinetic energy $E_\parallel = E_\text{kin} \cdot \cos^2{(\theta)}$ and the retarding potential $qU$\,\footnote{$qU$ is a positive quantity, since both the electron charge $q=-e$ and the retarding voltage are negative.}, see also eq. \ref{equ:model} later. This difference is called surplus energy $\Delta E_\text{s} = E_\parallel - qU$. In this case, the electron energy instead of the surplus energy is shown on the x-axis, to highlight the difference in absolute energy in the two measurements. The width of the transmission function gives an estimate on the energy spread of the produced electrons, in this case it is around \SI{127}{\milli\electronvolt} at \SI{32.2}{\kilo\electronvolt}.\\
 The performance of the upgraded source with this new design beyond the enhanced energy range will be discussed in the following section.

\section{Characteristics of produced photoelectrons\label{sec:expresponse}}

\begin{figure}
    \centering
    \includegraphics[width=0.6\linewidth]{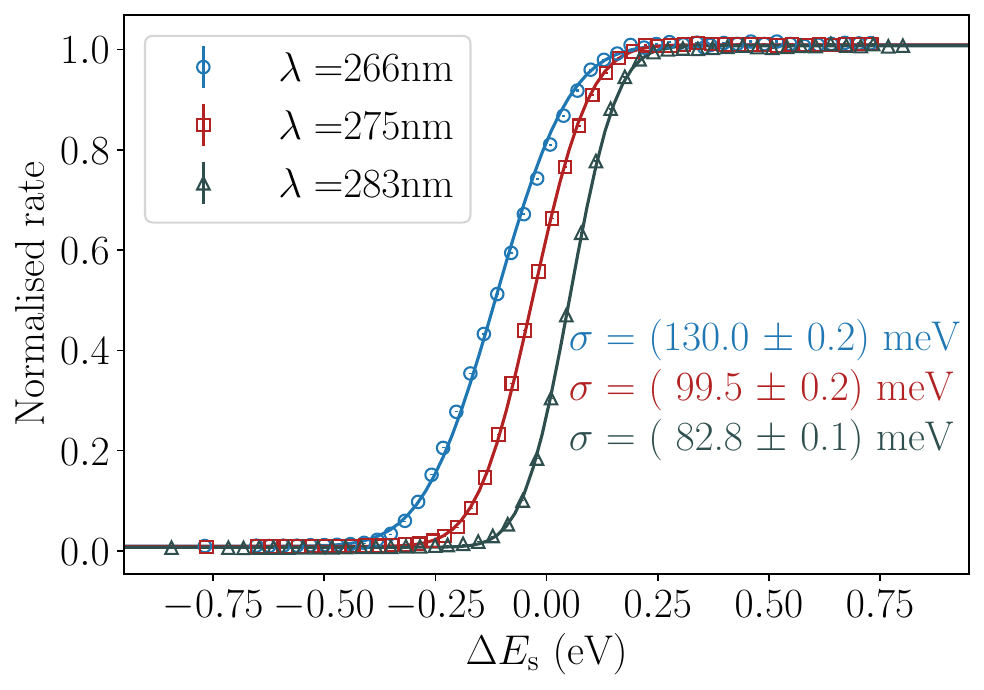}
    \caption{Transmission functions measured at different UV-light wavelengths. Shown is the normalized measured electron rate at the detector as function of the surplus energy $\Delta E_\text{s} = E_\perp - qU$, taken at a magnetic field in the analyzing plane of $B_\text{ana} = \SI{1}{G}$. The energy distribution gets broader for higher photon energies corresponding to lower wavelengths. The energy spread can be approximated as the width $\sigma$ of an error function fitted to the data, the obtained values for the three measurements are also shown in the figure.}
    \label{fig:tf-wavelength}
\end{figure}

\begin{figure}
    \centering
    \includegraphics[width=0.6\linewidth]{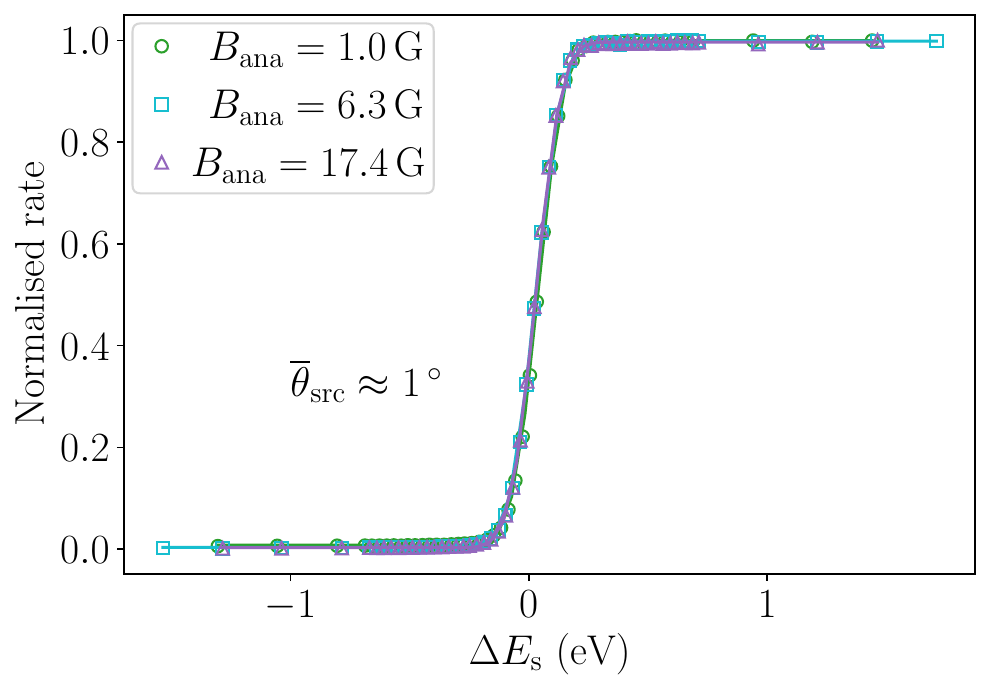}
    \includegraphics[width=0.6\linewidth]{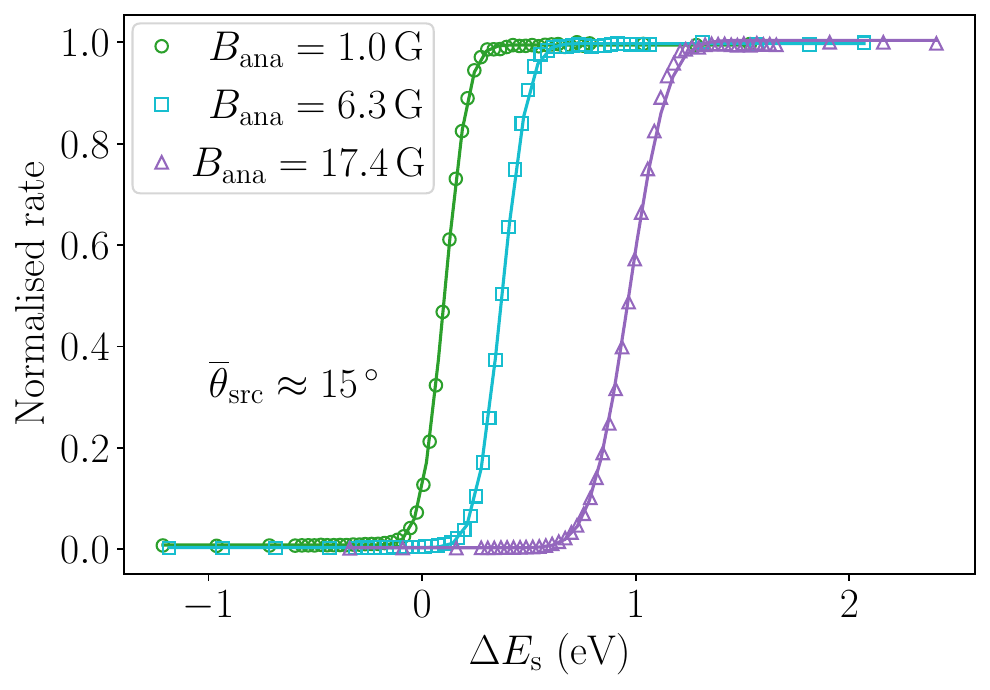}
    \caption{Transmission functions at \added{pitch angle} $\overline{\theta}_\text{src}\approx \SI{1}{\degree}$ (top) and $\overline{\theta}_\text{src}\approx\SI{15}{\degree}$ (bottom), measured at \SI{18.6}{\kilo\electronvolt} electron energy with different magnetic fields $B_\text{ana}$ in the analyzing plane. Clearly visible is the strong effect of $B_\text{ana}$ on the measured transmission functions at large mean pitch angle $\overline{\theta}$.}
    \label{fig:tf-bana}
\end{figure}

The key parameters of the photoelectron source are the energy distribution of produced photoelectrons\deleted{ on the one hand,} and their angular distribution\deleted{ on the other hand}. For calibration purposes, a narrow energy spread ($\ll \SI{1}{\electronvolt}$) is favorable. Regarding the angular distribution, a wide range of possible mean angles with an angular spread $\le \SI{1}{\degree}$ is desired. The characteristics of the upgraded photoelectron source regarding both key parameters are presented in the following.

\paragraph{Energy distribution}
 The ideal photoelectron source would produce purely mono-energetic electrons and therefore would have an energy distribution 
described by a $\updelta$-function. However, several effects lead to a smearing of the energy distribution:
\begin{itemize}
\item The non-zero operation temperature causes a smearing of the Fermi distribution within the metal itself and consequently a smearing of energies after emission.
\item The photon energy is not perfectly matching the cathode's mean work function, causing a mean energy of emitted electrons which is larger than zero.
\item The photocathode is not perfectly mono-crystalline and has an imperfect structure with non-vanishing surface roughness, a spatial change of layer thickness and adsorbants along the surface. This causes a spatial spread of the cathode's work function, leading to different energies of electrons emitted at different points of the cathode \cite{PhD-Behrens}.
\item The use of a light source with an imperfect wavelength selection - the LDLS light leaving the monochromator 
is expected to have a wavelength spread of $\qtyrange[range-units=bracket, range-phrase=-]{3}{10}{\nano\meter}$ - smears the photon energies. This effect does not occur when using the UV laser which has a sharp wavelength of $\SI{266}{\nano\meter}$.
\end{itemize}
The energy spread of the electrons depends on the light wavelength, \added{with} typical values \deleted{are }around \SI{100}{\milli\electronvolt}. In a measurement at low magnetic field in the analyzing plane, the energy spread can be observed from the width of the step-like transmission function. The analyzing plane is the vertical plane in the spectrometer at which the highest absolute retarding potential is reached. \replaced{Example}{Exemplary} functions measured with different UV-light wavelengths are shown in fig. \ref{fig:tf-wavelength}.

\paragraph{Angular distribution}
The energy distribution of the electrons produces a constant broadening of the transmission function which is independent from the spectrometer settings. In contrast, the effect of a non-zero angle $\theta$ of the produced electrons depends on the magnetic field in the analyzing plane $B_\mathrm{ana}$, because it defines the remaining \replaced{transverse}{transversal} momentum $p_\perp$ in the analyzing plane. The mean angle causes a shift, while the angular spread leads to a broadening of the transmission function. 
Making use of the conservation of the magnetic orbital momentum in case of adiabatic transport, see eq. \ref{equ:adiabatic_invariance} and eq. \ref{equ:relativistic_adiabatic_invariant}, and of the fact that the Lorentz factor in the analyzing plane is $\gamma_\text{ana} \approx 1$ due to the retardation, $E_{\perp,\text{ana}}$ is then given by
\begin{equation}
    E_{\perp,\text{ana}} = 
    \CWCW{E_{\perp,\text{src}} \cdot \frac{ B_\text{ana}}{B_\text{src}} \cdot \frac{\gamma_\text{src} + 1}{\gamma_\text{ana}+1} }= 
    E_\text{kin} \cdot \sin^2{(\theta_\text{src})} 
    \cdot \frac{B_\text{ana}}{B_\text{src}} \cdot \frac{\gamma_\text{src} + 1}{2}
\label{equ:shiftrel}
\end{equation}
with $B_\text{src}$, $\gamma_\text{src}$ and $\theta_\text{src}$ being the magnetic field strength, the Lorentz factor and the electron pitch angle in the tritium source, respectively.
Transmission functions measured at small and high mean pitch angle $\overline{\theta}$ with different magnetic fields $B_\text{ana}$ in the analyzing plane are compared in fig. \ref{fig:tf-bana}. A small pitch angle is chosen for measurements of the gas density in the source and measurements of energy-differential energy losses. Larger pitch angles can be used for systematic investigation of angular-dependent backscattering effects at detector and Rear Wall and for studies of adiabatic transport in the electromagnetic fields of the \mbox{KATRIN} spectrometer.

General expectations on the angular distribution after electron acceleration are obtained from simulations with the particle-tracking software \texttt{Kassiopeia}\,\cite{Furse_2017}, see \cite{Behrens2017}. However, because of the complexity of the electron transport within the Rear Section and the high impact of possible small misalignments of coils and electrodes during the strongly non-adiabatic acceleration process, a measurement of the produced electron pitch angles is more reliable. Here it should be considered that the minimum achievable pitch angle $\theta$ is not necessarily \SI{0}{\degree}, because the \Sch{electron source's electrodes} can only be tilted in one direction and a remaining non-zero angle \Sch{of the electrodes normal} to the magnetic-field lines can persist. Furthermore, an angle of \SI{0}{\degree} requires perfect alignment of electric and magnetic fields along the whole acceleration \Sch{region}, and also a finite angular spread \Sch{due to the emission process} will always lead to a non-zero mean angle. As $B_\text{src} = \SI{2.51}{\tesla} \gg B_\text{start} \approx \SI{25}{\milli\tesla}$, a small angle $\theta_\text{1} = \theta_\text{start}$ at the electron source, at which the magnetic field is $B_\text{start}$, will lead to angles $\theta_\text{2} = \theta_\text{src}$ significantly larger than \SI{0}{\degree} in the tritium source according to eq. \ref{eq:adiabatic_transport}. 

The uncertainty on $\theta$ directly propagates into the uncertainty on the gas density: The gas density is determined by measuring the mean scattering probability $\mu_\text{\added{scat}}$ of electrons emitted from the photoelectron source, as $\mu$ and the gas density integrated along the path length through the source, $\rho d$, are related via
\begin{equation}
    \rho d = \frac{\mu_\text{\added{scat}}}{\sigma_\text{inel,tot}},
\end{equation}
with the total inelastic cross-section of electrons in tritium gas $\sigma_\text{inel,tot}$. The path length is increased for electrons with non-zero angle $\theta_\text{src}$ at magnetic-field strength $B_\text{src}$, leading to an effective path length $d_\text{eff}$, such that\,\cite{Kleesiek:2018mel}
\begin{equation}
    \rho d_\text{eff} = \frac{\rho d}{\cos{\theta_\text{src}}}\,.
\end{equation}
The contribution of the angle uncertainty $\delta \theta$ on the gas-density uncertainty is therefore $\propto \tan{\theta}\cos{\theta}^{-1}\delta \theta$ and increases with $\theta$ for $\SI{0}{\degree} < \theta < \SI{45}{\degree}$.
A precise knowledge on the mean pitch angle $\overline{\theta}$ in the tritium source is therefore required, and the choice of a small angle for gas-density measurements is advantageous. Information on $\theta$ for the photoelectrons can be obtained from the measured transmission function of the electron source.\\ 
\begin{figure}
    \centering
    \includegraphics[width=0.6\linewidth]{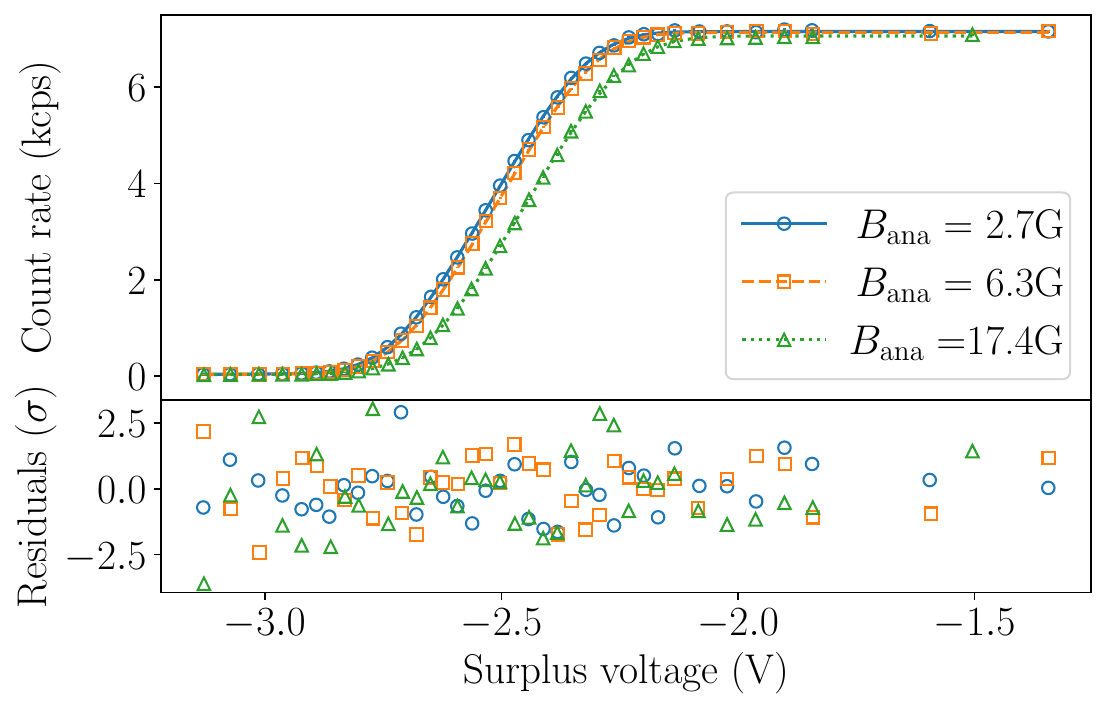}
    \caption{Combined fit of transmission functions measured at \SI{18.8}{\kilo\electronvolt} electron energy with different $B_\text{ana}$ using the extended model description of eq. \ref{equ:model} to eq. \ref{equ:eta}. Shown on the x-axis is the surplus voltage $U - U_\text{back}$, which is negative when the retarding voltage $U$ is more negative than the voltage $U_\text{back}$ on the backplate of the photoelectron source. The obtained parameters for the angular distribution $\zeta(\theta)$ are $\hat{\theta} = \SI[separate-uncertainty=true]{5.0\pm 0.3}{\degree}$ (mean angle), $\sigma_\theta = \SI[separate-uncertainty=true]{2.0\pm 0.6}{\degree}$ (angular spread) and $\tau_\theta = {1.8\pm 0.4}$ (shape parameter). The weighted mean of the fitted $\zeta(\theta)$ is $\bar{\theta} = \SI{4.7}{\degree}$.}
    \label{fig:modelfit}
\end{figure}

The measured transmission function $T(E, qU, qU_\text{back})$ for a certain electron energy $E$, a retarding voltage $qU$ and the acceleration voltage $U_\text{back}$ of the photoelectron source depends on the energy distribution $\eta(E, qU_\text{back})$ and the angular distribution $\zeta(\theta)$ of the produced electrons. Analytically, it can be described via\,\cite{Behrens2017,PhD-Behrens,Erhard_2016}
\begin{equation}
    T(E, qU, qU_\text{back}) = R_\text{s}\cdot \int_E^\infty \eta(\epsilon, qU_\text{back}) \int_0^{\theta_\text{max}(\epsilon, qU, B_\text{ana})}\zeta(\theta)d\theta d\epsilon\, + R_\text{b},
    \label{equ:model}
\end{equation}
where 
\begin{equation} 
\theta_\text{max}(\epsilon, qU, B_\text{ana}) =  \left\{
\begin{array}{ll}
\min{\left(\arcsin{\left(\sqrt{ \frac{\epsilon - qU}{\epsilon} \frac{2}{1+\gamma} \frac{B_\text{src}}{B_\text{ana}}}\right)}, \,\arcsin{\left(\sqrt{\frac{B_\text{src}}{B_\text{max}}}\right)}\right)} & , \frac{\epsilon - qU}{\epsilon} \ge 0\\
0 & \text{, else} \\
\end{array}
\right.
\end{equation} 
is the maximum transmitted pitch angle for a certain $E, qU$ and magnetic field $B_\text{ana}$ in the analyzing plane, $R_\text{s}$ is the signal rate and $R_\text{b}$ the background rate. $B_\text{max} = \SI{4.24}{\tesla}$ is the maximum magnetic field along the whole KATRIN beamline, it limits the maximum transmitted angle due to magnetic reflection. $\eta(E, qU_\text{back})$ is expected to be asymmetric due to the Fermi distribution, and a generalized normal distribution to describe $\eta (E, qU_\text{back})$ was introduced in \cite{Behrens2017}. The angular distribution was described by a sum of two normal distributions, to take into account the shape deformation at $\theta \rightarrow \SI{0}{\degree}$ caused by the fact that negative polar angles do not occur. While these representations were sufficient to describe data from the first Rear-Section electron source, the higher rate and correspondingly higher statistics gained in measurements with the upgraded electron source require an extension of the model. First, the model of the energy distribution is extended by another generalized normal distribution, such that
\begin{equation}
    \eta(E, qU_\text{back}) = (1-w)\cdot \eta_1(E - qU_\text{back}) + w \cdot \eta_2(E - qU_\text{back}) 
\end{equation}
with
\begin{equation}
\eta_k(E_\text{e}) = \frac{1}{\sqrt{2\pi}}\cdot 
\left\{ \begin{array}{ll}
 \frac{1}{\sigma_{\text{E},k}} \cdot \exp{\left( -\frac{1}{2}(\frac{E_\text{e}-\hat E_k}{\sigma_{\text{E},k}})^2\right)}, &\tau_{\text{E},k} = 0\\ 
\frac{1}{\sigma_{\text{E},k} - \tau_{\text{E},k}(E_\text{e}-\hat E_k)}\cdot \exp{\left( -\frac{1}{2\tau_{\text{E},k}^2}\cdot \ln{\left(1 - \tau_{\text{E},k}\frac{E_\text{e}-\hat E_k}{\sigma_{\text{E},k}}\right)^2}\right)}, &\tau_{\text{E},k}\neq 0 \\
\end{array} \right.\ ,
\label{equ:eta}
\end{equation}
where $k = {1, 2}$, $\tau_{\text{E},k}$ is a shape parameter and $w\in [0, 1]$ a weighting parameter. $E_\text{e} = E - qU_\text{back}$ is the photoelectron energy after emission, thus before acceleration. The extension can be justified by the
fact that the imperfect surface of the photocathode has more than one work function and correspondingly more than one energy distribution. Second, the description of the angular distribution is changed to a generalized normal distribution. 

To disentangle the impact of the different parameters describing energy and angular distribution, a combined fit of transmission functions measured with the same configuration of the electron source, but with differing magnitudes of $B_\text{ana}$ can be performed, making use of the differing impact of the angular distribution at different $B_\text{ana}$. An exemplary fit to data taken with the upgraded electron source is shown in fig. \ref{fig:modelfit}. The fit yields $\chi^2/ndof = 154.6/95=1.63$, which is significantly larger than 1, but without clear structures in the residuals. However, the computational fitting effort is tremendous due to the large number of free parameters, and the fit result suffers from a high correlation between the various fit parameters. Therefore, a simpler approach, which does not require detailed knowledge on the exact energy and angular distribution, is introduced in the following.

\section{Estimation of electron pitch angle from transmission-function measurements \label{sec:pitchangle}}

\begin{figure}
    \centering
        \includegraphics[width=0.6\linewidth]{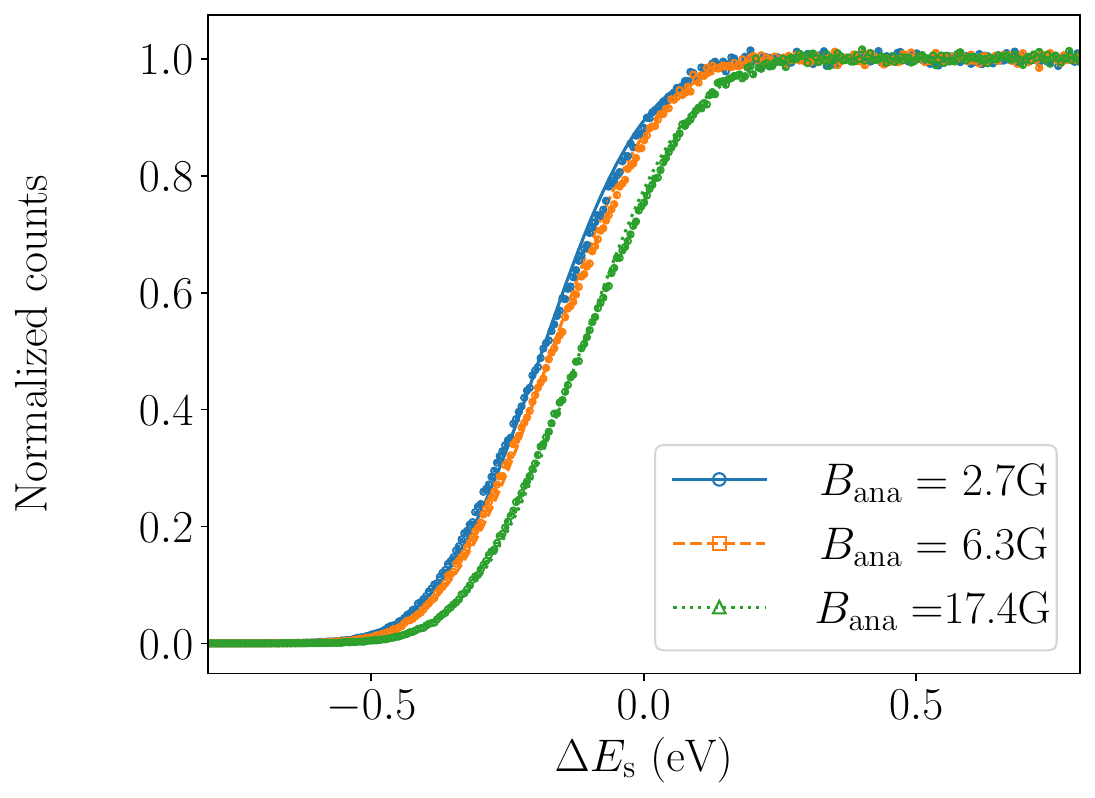}
      \includegraphics[width=0.6\linewidth]{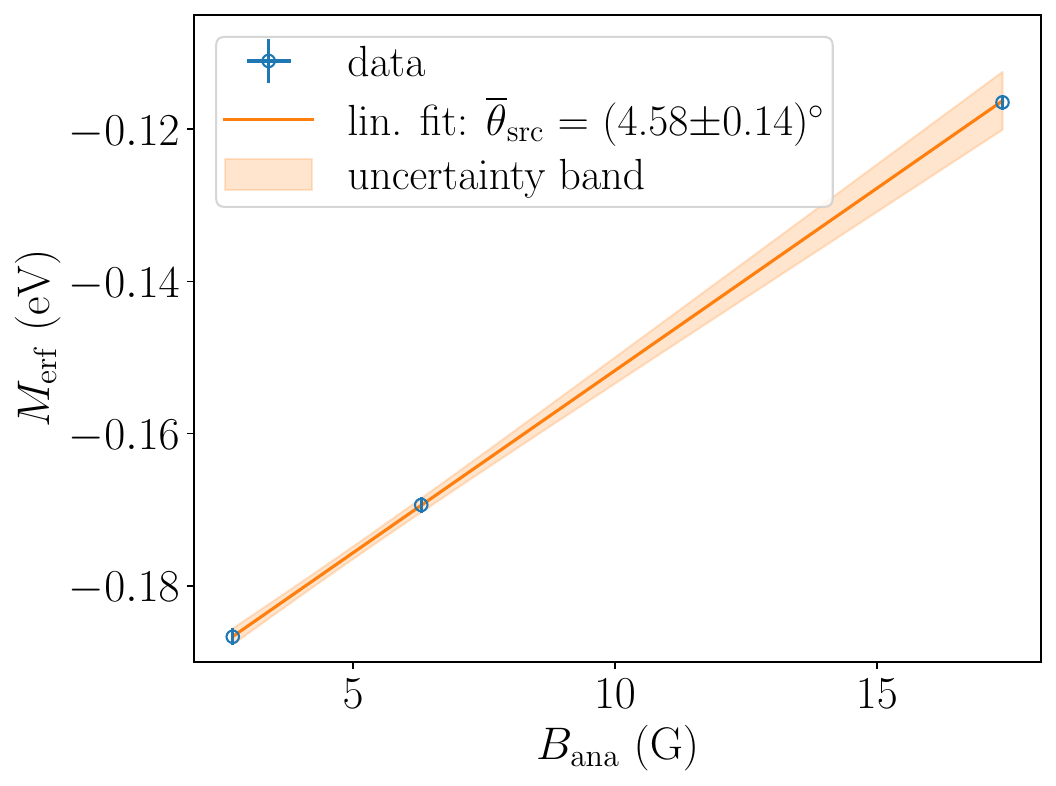}
    \caption{Determination of the electron pitch angle from simulated transmission-function scans at three different $B_\text{ana}$. Top: Monte-Carlo data, generated according to the model described by eq. \ref{equ:model} to eq. \ref{equ:eta} and using input distributions which are representative for the upgraded photoelectron source with $\overline{\theta}_\text{MC} =~$\replaced{\SI{4.42}{\degree}}{$4.4\,^\circ$}. For each magnetic field, $10^7$ events were generated. The lines show error-function fits applied to the data. Bottom: Determination of $\overline{\theta}$ from the linear relation between the middles of transmission $\added{M}_\text{erf}$ and $B_\text{ana}$. The uncertainties on $\added{M}_\text{erf}$ are scaled by the reduced $\chi^2$ values from the error-function fits.}
    \label{fig:method-verification}
\end{figure}

A robust method to determine the mean pitch angle without exact knowledge of the energy and angular distribution is the relative comparison of the average surplus-energy threshold at which electrons are transmitted, the middle of transmission\footnote{It corresponds to the surplus energy at which \SI{50}{\percent} of the electrons are transmitted.}, determined at different $B_\mathrm{ana}$.
The shift of the middle of transmission due to a non-zero mean angle 
$\overline{\theta}$ of electrons with kinetic energy $E_\text{kin}$ has its
origin in the remaining \replaced{transverse}{transversal} energy $E_\perp(B_\text{ana})$ in the analyzing plane, caused by the finite energy resolution of the spectrometer. The \replaced{transverse}{transversal} energy $E_\perp$ can be derived from the remaining \replaced{transverse}{transversal} momentum $p_\perp$ in the analyzing plane corresponding to a certain pitch angle \added{$\theta_\text{src}$} according to eq. \ref{equ:shiftrel}.
As the Lorentz factor for  electrons with $E_\text{kin} = \SI{18.6}{\kilo\electronvolt}$ is $\gamma_\text{src} = 1.04\approx 1$, it can be simplified to 
\begin{equation}
    E_{\perp,\text{ana}} = \underbrace{\frac{E_\text{kin}}{B_{\text{src}}} \cdot \text{sin}^2(\theta_\text{src})}_{\text{slope}~s} 
    \cdot B_{\text{ana}}\,.
\label{equ:shift}
\end{equation}
\added{According to eq. \ref{equ:shift}, the transverse energy $E_{\perp,\text{ana}}$ depends on the sine squared of the pitch angle at the source $\sin^2(\theta_\mathrm{src})$ and the magnetic field in the analyzing plane $B_\mathrm{ana}$, which allows to determine the pitch angles by measuring the transmission function of different magnetic fields in the analyzing plane.}
To obtain the middle of transmission from a measurement, the experimental response can be approximated by an error function. The error-function description is not physically motivated, but is only an effective description. In particular, it is not able to differentiate between the energy distribution and the angular distribution of the electrons - the smearing caused by both distributions is effectively described in a single smearing by the $\sigma$-parameter of the error function. However, the angular spread is small, and the energy spread is the dominant contribution to the total broadening of the transmission function. Thus,
the approximation via an error function mainly leads to a constant shift of the middle of transmission determined from the mean position of the error function, and the relative shift for different $B_\text{ana}$ can be neglected. The obtained middles of transmission as function of $B_\text{ana}$ follow a linear relation, of which the slope is \footnote{The $\gamma$-factor can also easily be included in eq. \ref{equ:slope}, which is reasonable especially at higher electron energies. At \SI{18.6}{\kilo\electronvolt}, the neglect of the $\gamma$-factor causes a bias of \SI{1}{\percent} on the obtained angle.}
$s=\frac{E_\text{kin}}{B_{\text{src}}} \cdot \text{sin}^2(\theta_\text{src})$. Thus, the mean electron angle in the tritium source can be determined via
\begin{equation}
    \overline{\theta}_\text{src} = \arcsin{\left(\sqrt{\frac{s\cdot B_\text{src}}{E_\text{kin}}}\right)}.
    \label{equ:slope}
\end{equation}
This method \replaced{becomes}{is} more sensitive\replaced{ as}{, the higher} the difference between minimal and maximal $B_\text{ana}$ \replaced{grows}{is}. At \mbox{KATRIN}, the used values of $B_\text{ana}$ vary between \SI{2.7}{G} and \SI{17.4}{G}, which are the currently technically feasible field limits in which a well-defined analyzing plane can be obtained. In the analyzing plane, the real retarding potential is altered by the so-called potential depression compared to the voltage set point applied to the spectrometer. The potential depression differs for different magnetic-field settings. This needs to be considered when comparing the middles of transmission. The potential depression is simulated with the \mbox{KATRIN}-specific field and particle-tracking software \texttt{Kassiopeia}. \added{It is $\SI{2.34}{\electronvolt}$ for all used magnetic-field settings, the depression change between the settings is a few meV. The measured middles of transmission are shifted by the setting-dependent potential depression in the analysis.}

This analysis procedure for angle determination was verified via a Monte-Carlo simulation. Transmission functions at different $B_\text{ana}$ were simulated using the model described by eq. \ref{equ:model} and eq. \ref{equ:eta}, see top of fig. \ref{fig:method-verification}. The simulated functions were then fitted with error functions, which are shown as lines in the figure. The obtained mean positions $\added{M}_\text{erf}$ of the error functions as function of $B_\text{ana}$ are shown in fig. \ref{fig:method-verification} bottom. The fitted slope yields $\overline{\theta} = \SI[separate-uncertainty=true]{4.58\pm 0.14}{\degree}$ (statistical uncertainty only), in good agreement with the Monte-Carlo input $\overline{\theta}_\text{MC} = \SI{4.42}{\degree}$. Applying the described procedure to the data from fig. \ref{fig:modelfit} yields $\overline{\theta}_\text{erf} = \SI[separate-uncertainty=true]{4.91\pm 0.22}{\degree}$, which is compatible with the weighted mean $\overline{\theta} = \SI{4.7}{\degree}$ obtained from the fit result using the transmission-function model from sec. \ref{sec:expresponse}.

Overall, the presented method provides a quick and reliable way of for angle determination from transmission-function measurements at different $B_\text{ana}$, with a required measurement time of around \SI{1.5}{\hour} and negligible analysis effort. Since 2022, this characterization measurement has been performed several times per year to continuously monitor the electron angle of the photoelectrons.\\\\
\deleted{After installation of the upgraded electron source, an enhanced background level compared to the predecessor was observed. To resolve this, a novel hardware-based option for background reduction in the pulsed mode was developed, which is introduced in the following.}

\section{Background reduction by a pulsing approach \label{sec:background}}
\added{After installation of the upgraded electron source, an enhanced background level compared to the predecessor was observed. To resolve this, a novel hardware-based option for background reduction in the pulsed mode was developed.}

In addition to photoelectrons (termed "signal"), the electron source also produces background electrons with a different production mechanism. Tritium ions created in the vicinity of the source are accelerated by the negative high voltages applied to the electrodes towards the backplate, where they sputter off electrons that are subsequently accelerated to similar energies as the signal electrons\,\cite{KATRIN:2021rqj}. In the pulsed measurement mode, signal electrons are emitted in time intervals of $\mathcal{O}(\SI{10}{\nano\second})$, which repeat every \SI{50}{\micro\second}. In the time window between two intervals, only background electrons are emitted. A background reduction is therefore possible by preventing electrons from passing the opening in the Rear Wall in between signal bunches. 

Electrons can only pass the whole Rear Section if the currents and voltages of all coils and electrodes from fig. \ref{fig:rearsection} are chosen such that the beam deflection allows the electron beam to pass both the chicane behind the dipole electrodes and the Rear-Wall hole. This can be utilized by changing the dipole-electrode setting synchronous to the laser-trigger signal between two states: One state in which electrons \Sch{pass} the Rear-Wall hole, and a second state in which they are blocked on their way through the Rear Section.

In the standard measurement configuration, electrons are only transmitted
to the detector if the dipole electrodes are set to $\pm\SI{400}{\volt}$. If one dipole electrode is set to $+\SI{400}{\volt}$ and the other one to $\SI{0}{\volt}$, transmission is prevented. Background reduction is therefore possible by pulsing one dipole electrode between $-\SI{400}{\volt}$
and $\SI{0}{\volt}$. The capacitance $C$ of the electrode was measured to be around \SI{570}{\pico\farad}. With $U = \SI{-400}{\volt}$, the charge on the electrode is $Q = C\cdot U = \SI{2e-7}{C}$. To transfer this charge within a time frame of $t \approx \mathcal{O}(\SI{100}{\nano\second})$, a current of $I = Q/t = \SI{2.3}{\ampere}$ is required.

To realize a rectangular voltage pulse with \SI{400}{\volt} amplitude and sufficient slew rate at this capacitive load, a circuit with a RECOM\footnote{RECOM Power GmbH $\&$ RECOM Engineering GmbH $\&$ Co KG, Münzfeld 35, 4810 Gmunden, Austria} 
R-REF01-HB MOSFET half-bridge board is used. The board features a 
maximum voltage slope of \SI{65}{\kilo\volt\per\micro\second} for voltages up to \SI{1000}{\volt}, and a maximum gate-drive current of \SI{10}{\ampere}.
The low-voltage trigger signal of the bridge is synchronized with the laser-trigger signal. A time delay between the two signals accounts for time differences between electron emission, trigger signal and half-bridge switching.\\
\begin{figure}
    \centering
    \includegraphics[width=0.6\linewidth]{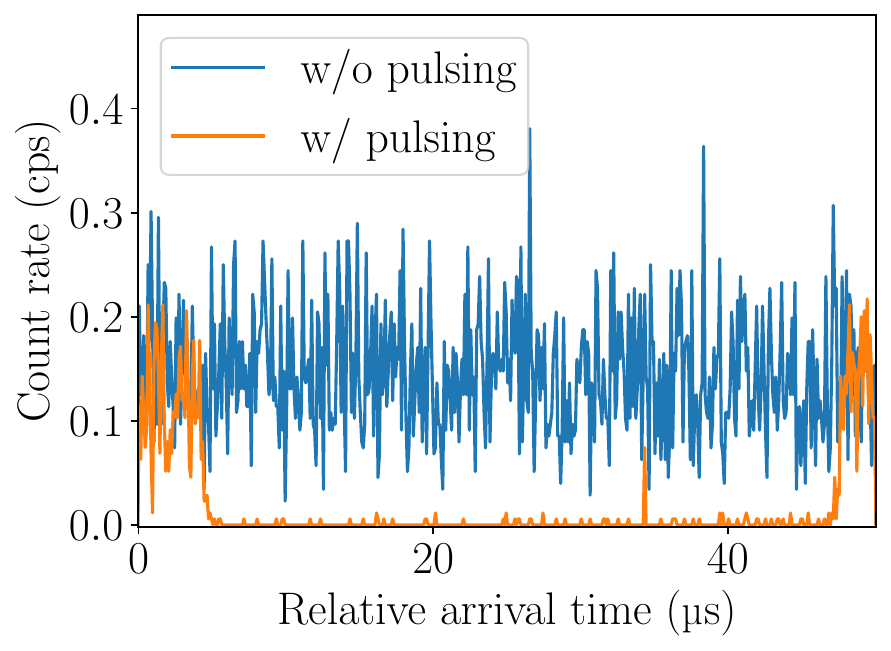}
    \includegraphics[width=0.6\linewidth]{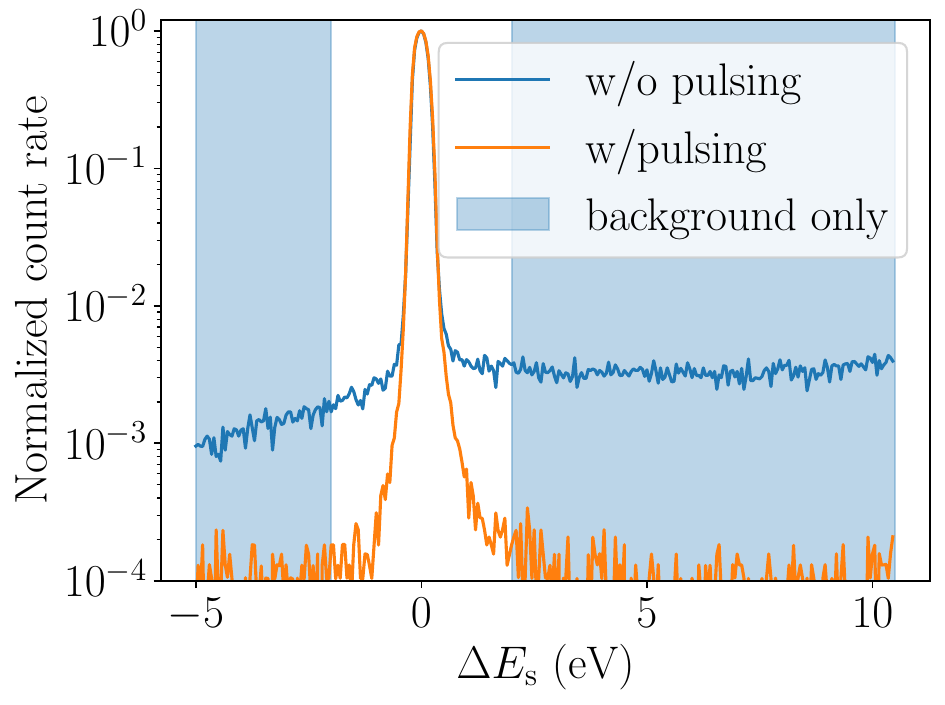}
    \caption{Impact of dipole-electrode pulsing on background rate.
    Top: Arrival-time distribution of the background. Without pulsing, the background arrives uniformly distributed in time. With pulsing, background occurs only at arrival times above \SI{47}{\micro\second} and below \SI{5}{\micro\second}. 
     Bottom: \added{Energy-d}ifferential signal of the electron source with and without pulsing. Shown are the measured counts, normalized to their maximum at \SI{0}{\electronvolt} surplus energy, as function of the surplus energy $\Delta E_\text{s}$. Especially in the background-only region at surplus energies between \SI{2}{\electronvolt} and \SI{10}{\electronvolt}, in which no signal electrons are measured due to the minimum inelastic energy loss of $\approx \SI{11}{\electronvolt}$\,\cite{KATRIN:2021rqj}, the effect of background reduction is clearly visible.}
    \label{fig:pulsing}
\end{figure}

 A comparison of the arrival-time distribution at the detector of background measured with and without dipole-electrode pulsing is shown in fig. \ref{fig:pulsing} top. As expected, without active pulsing, the background arrives uniformly in time. With active pulsing, background is only measured in a time window corresponding to the time interval in which the pulsed electrode is at \SI{-400}{\volt} and electrons can be transmitted. This interval has a length of around \SI{7}{\micro\second}. Due to the reduced time window in which the background is transmitted, a background reduction by a factor of 7, from \SI{73}{cps} without pulsing to \SI{9}{cps} with pulsing at \SI{200}{\electronvolt} surplus energy, is achieved. This makes the background level of the new source compatible to the old source's one. A comparison of the \added{energy-}differential \added{electron} signal with and without active background reduction is shown in fig. 
\ref{fig:pulsing} bottom. \replaced{The energy-differential signal is obtained by combining the high-pass filter of the spectrometer with a low-pass filter from the flight-time information of the electrons.}{The differential signal is obtained as follows: In addition to the spectrometer, which acts as high-pass filter for the electron energy, also the flight time of the electrons is used.} The higher the surplus energy of the electrons emitted in a signal bunch, the earlier they arrive at the detector. Considering only electrons above a certain arrival-time threshold therefore acts as \deleted{an additional }low-pass filter for the energy. In the standard measurement without active pulsing, the low-pass filter does not work for the background electrons, as their arrival time is random and independent on their surplus energy. Essentially, an integral background spectrum instead of an energy-differential one is measured, which is why background is observed along all surplus energies \added{in the blue curve of fig. \ref{fig:pulsing} bottom}. In contrast, with active pulsing \replaced{the background electrons have a non-uniform arrival-time distribution}{also the background electrons receive a distinct arrival time} at the detector.\deleted{, which makes the energy low-pass filter using the flight time working also for the background electrons.}
Since only electrons above a certain flight time are considered in this so-called time-of-flight method\,\cite{KATRIN:2021rqj}, the background electrons occurring at arrival times below \SI{20}{\micro\second} in fig. \ref{fig:pulsing} top are not counted at all, which leads to a background reduction in the time-of-flight mode by \deleted{even }a factor of 10.

\section{Conclusions}
An upgrade of the calibration photoelectron source at the \mbox{KATRIN} Rear Section was presented. The source features a high-statistics electron beam with $\mathcal{O}(\SI{20}{kcps})$ rate, compared to $\mathcal{O}(\SI{1}{kcps})$ with the previous source. It reaches electron energies up to \SI{32}{\kilo\electronvolt}, with an energy spread between \SI{80}{\milli\volt} and \SI{130}{\milli\volt}, while the predecessor could only provide energies up to \SI{20}{\kilo\electronvolt}. It allows for a precise adjustment of the electron pitch angle with regard to the magnetic-field lines, with possible mean angles between $\SI{1}{\degree}$ and $\mathcal{O}(\SI{40}{\degree})$ and typical angular spreads of $\mathcal{O}(\SI{1}{\degree})$ at a magnetic field strength of \SI{2.51}{\tesla}. The minimum angle achieved with the earlier source was around \SI{10}{\degree} and a reproducible adjustment of the angle was not possible. 

\added{While the current neutrino-mass limit of \SI{0.45}{\electronvolt\per c^2} set by KATRIN\,\cite{KATRIN:2024cdt} is dominated by the statistical uncertainty, a very good knowledge on the systematic uncertainties is required for the analysis of the final KATRIN data set with more than a factor 5 higher statistics. The dominant systematic contributions are energy-losses of $\upbeta$-decay electrons due to scattering in the tritium source as well as the gas density in the tritium source. Both of these inputs are determined experimentally by means of measurements with the photoelectron source.}
\replaced{The new source features allow not only for an even more accurate determination of the energy-loss function than before and the frequent precise determination of the gas density in more than two thirds of the KATRIN neutrino-mass campaigns, but also for an in-situ measurement of}{The new features allow for an accurate determination of experimental parameters like the energy-loss function of $\upbeta$-decay electrons in the tritium source, the gas density in the tritium source and} angular-dependent effects such as electron backscattering at the detector and at the gold-plated disk at the rear end of the tritium source in the \mbox{KATRIN} environment.

\replaced{As a very precise knowledge on the electron pitch angle of the calibration electrons is required, a}{A} new technique for an experimental determination of the electron pitch angle \replaced{was developed. It makes}{, making} use of the field-dependent sensitivity of the \mbox{KATRIN} spectrometer to non-zero electron angles\replaced{ and}{,} delivers knowledge on the pitch angle with $< \SI{1}{\degree}$ precision at \SI{2.51}{\tesla} field strength. In addition, the background level of the electron source, typically being of $\mathcal{O}(\SI{70}{cps})$, was shown to be reduced by at least a factor of 7 in the pulsed operation mode, using a periodic electron deflection by $E\times B$ drift induced by dipole electrodes. The lower background level in particular allows for improved measurements of \added{the} energy losses of electrons due to scattering off tritium molecules in the tritium source, the results of which will be published elsewhere.

\section*{Acknowledgments}
We acknowledge the support of \replaced{Federal Ministery of Research, Technology and Space}{Ministry for Education and Research} BMFTR (contract number 05A20PMA, 05A23PMA) and Deutsche Forschungsgemeinschaft DFG (Research Training Group GRK 2149) in Germany. We kindly thank Caroline Rodenbeck for her helpful contributions in the electron-source related discussions, and Armen Beglarian for his support during the installation of the electron source in the KATRIN beamline.

\section*{Declarations}
\paragraph{Funding}
This work was supported by the German Ministry for Education and Research BM\CWCW{FTR} (05A20PMA, 05A23PMA) and Deutsche Forschungs\-gemeinschaft DFG (Research Training Group GRK 2149).
\paragraph{Conflict of interest/Competing interests}
The authors have no relevant financial or non-financial interests to disclose.
\paragraph{Availability of data and materials}
The datasets generated during and/or analyzed during the current study are available from the corresponding author on reasonable request.
\paragraph{Code availability}
Not available
\noindent
\bibliography{bibliography.bib}

\end{document}